# STILN: A Novel Spatial-Temporal Information Learning Network for EEG-based Emotion Recognition

Yiheng Tang [a], Yongxiong Wang [a,*], Xiaoli Zhang [a], Zhe Wang [a]

[a] *The School of Optical-Electrical and Computer Engineering, University of Shanghai for Science and Technology, Shanghai 200093, China.*

**ABSTRACT**

The spatial correlations and the temporal contexts are indispensable in Electroencephalogram (EEG)-based emotion recognition. However, the learning of complex spatial correlations among several channels is a challenging problem. Besides, the temporal contexts learning is beneficial to emphasize the critical EEG frames because the subjects only reach the prospective emotion during part of stimuli. Hence, we propose a novel Spatial-Temporal Information Learning Network (STILN) to extract the discriminative features by capturing the spatial correlations and temporal contexts. Specifically, the generated 2D power topographic maps capture the dependencies among electrodes, and they are fed to the CNN-based spatial feature extraction network. Furthermore, Convolutional Block Attention Module (CBAM) recalibrates the weights of power topographic maps to emphasize the crucial brain regions and frequency bands. Meanwhile, Batch Normalizations (BNs) and Instance Normalizations (INs) are appropriately combined to relieve the individual differences. In the temporal contexts learning, we adopt the Bidirectional Long Short-Term Memory Network (Bi-LSTM) network to capture the dependencies among the EEG frames. To validate the effectiveness of the proposed method, subject-independent experiments are conducted on the public DEAP dataset. The proposed method has achieved the outstanding performance, and the accuracies of arousal and valence classification have reached 0.6831 and 0.6752 respectively.

**Keywords** - Emotion recognition, Electroencephalogram (EEG), Spatial correlations, Temporal contexts, Attention mechanism

## 1. Introduction

The research of brain and cognitive science aims to explore the linkage between human psychology and brain[1][2]. EEG possesses the advantages of easy acquisition and accurate reflection of emotional state in physiological signals [3]. The purpose of the EEG-based emotion recognition is to build harmonious human-computer interaction and endow the system with the ability to distinguish and comprehend the emotions. Therefore, EEG-based emotion recognition becomes an important research direction in cognitive neuroscience and computer science.

Psychologists establish discrete and dimensional models to quantify the emotional state. In the discrete emotion model, the emotional states can be divided into variety of discrete basic elements, such as happiness, fear, sadness, disgust, etc. [4]. The Valence-arousal plane[5] proposed by J. Russel is an extensively used dimensional emotion model which maps emotions into a 2D space, as shown in Fig.1(a). Arousal represents emotional state of calm to excitement. Valence refers to positive or negative mental activity. Self-assessment Manikins (SAM) System[6] is an effective emotion annotation method and it could enhance the consistency of the ratings among different participants. The ratings of each dimension are distributed from 1 to 9. Based on the valence-arousal space and SAM System, the emotional state could be more precisely represented.

---

* corresponding authors.
E-mail addresses: wyxiong@usst.edu.com (Y.-X. Wang)

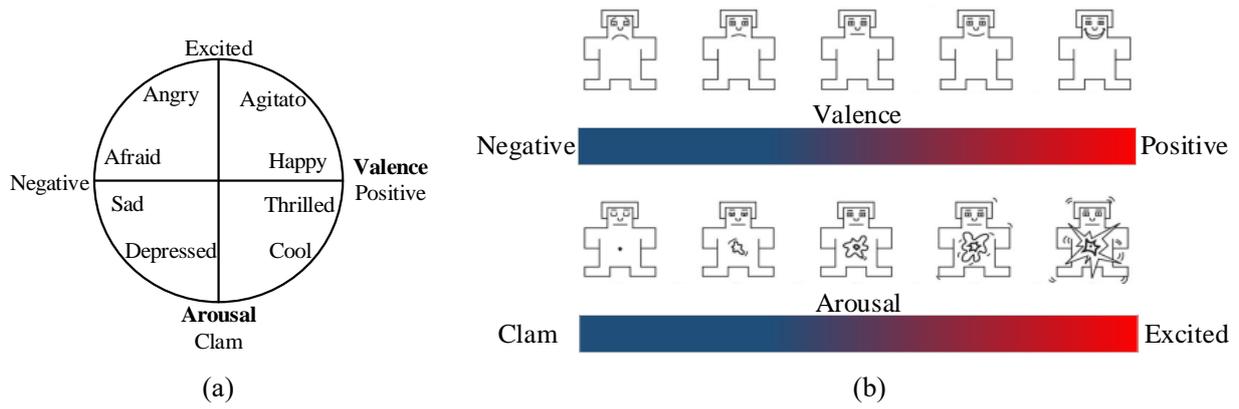

Fig.1 Emotion quantitation methods (a) The valence-arousal space (b) The cartoon representation models of the SAM system

The intrinsic attributes of EEG are multichannel, nonlinear, and nonstationary. The spatial correlations capture the dependencies among electrodes, and the nonlinear and nonstationary of EEG is reflected in time-domain. Therefore, EEG-based feature learning can be summarized as three points: temporal feature learning, spatial correlations learning, and multiple domain feature learning.

A. Temporal Feature Learning

In recent studies, EEG temporal domain features receive increasing interest from researchers. Badicu *et al.* [7] propose a hierarchical structure based on the long short-term memory network (LSTM), the LSTM layer is utilized to model time/feature sequences and to make predictions. Li *et al.* [8] extract power spectral density (PSD) features through time-frequency analysis, and they utilize LSTM to capture the temporal dynamics of emotions. Hwang *et al.* [9] extract EEG differential entropy features and employ LSTM to learn temporal contexts. Undeniably, the prospective emotion of the subjects can only be achieved in part of the stimulation process, the learning of the temporal contexts is profitable to emphasize the critical EEG frames. Their method verifies the necessity of the temporal contexts. However, these studies concentrate on surveying the fluctuation of EEG, but they do not make full use of the correlations among the electrodes.

B. Spatial Correlations Learning

EEG spatial information provides electrodes-correlated features which is beneficial to discriminate emotions. Khosrowabadi *et al.* [10] propose the feedforward neural network method, and enhanced the connectivity between different brain regions. However, the method deficiently captures the correlations between multi-electrode channels. Jatupaiboon *et al.* [11] utilized the support vector machine (SVM) to learn the asymmetric brain activity in different emotional states. Bashivan *et al.* [12] generate spectral images where EEG activities are converted into a series of topological structures. Then, the feature extractor is used to robustly learn the spatial correlations of EEG channels. Moon *et al.* [13] transformed the phase locking value connectivity features into the 2D matrix and adopted CNN to learn the spatial correlations. These methods prompt the development of EEG-based emotion recognition, but the dependence between electrodes has inadequately studied. Moreover, the neuroscience researches also demonstrate that human emotion production associate with the frontal lobe of the brain [14], the region that provides more contributory information than other regions. However, studies such as[10]-[13] have insufficient ability to emphasize information which is beneficial to discriminate emotions from several electrodes or brain regions.

C. Multiple Domain Feature Learning

The aforementioned methods commonly perform contexts learning only in the temporal domain, or only spatial correlations learning is integrated into the emotion recognition model. The fusion of spatial-temporal features ensures that more EEG features are extracted, so many researchers focus on the spatial-temporal fusion strategies. Guo *et al.* [15] fed the correlation coefficient matrix and synchronous likelihood matrix to the feature fusion framework of the emotional network and the inception network is adopted to fuse the latent features. Liu *et al.* [16] proposed a 3D convolution attention neural network composed of spatial-temporal feature extraction module and channel attention weight learning module, and the internal spatial correlations of multi-channel EEG signals during continuous period time are extracted. Wang *et al.* [17] propose the spatial-temporal feature fusion network to extract classification features and integrate feature complementary relationships. The multi-layer perceptron (MLP) is applied to learn and extract temporal features, and spatial-temporal feature fusion by Bi-LSTM. Although the spatial-temporal fusion strategies [15]-[17] adopt abundant information to improve the performance, it is inadequate learning for temporal contexts which is also discriminative to the emotional states.

As aforementioned, the following two issues can be identified: 1) The learning of complex spatial correlations among the electrodes have been inadequately studied, the essential electrodes and brain regions are necessary to be emphasized. 2) The EEG temporal contexts learning is effective way to capture the dependencies among the EEG frames and discriminate the emotions. Thus, we propose a spatial-temporal information learning network to extract discriminative features required for emotion recognition by capturing spatial correlations and temporal contexts robustly. The STILN model encompasses the following two elements:

(1) Spatial Correlations Learning

We use the frequency bands characteristics of windows in EEG and the position of electrodes to generate the 2D power topographic maps of PSD features[17]. Different channels or different spaces of the feature maps contribute variously to the classification accuracy, and are not fully investigated. In the STILN model, CBAM recalibrates the weights of channels and space in the power topographic maps, the crucial brain regions and frequency bands are emphasized, and that is one of the differences in our works as compared with other models. BNs and INs are exploited to preserve the underlying features of EEG signal while mitigating the individual differences. The spatial correlations of the EEG are learned by the extraction network and the fusion network, and it greatly enhances the accuracy of emotion recognition.

(2) Temporal Contexts Learning

We adopt a more robustly temporal feature learning methods, the temporal contexts learning, rather than the spatial-temporal fusion strategy. Hence, Bi-LSTM is applied to bidirectional learning the temporal contexts of different EEG windows, and it captures the essential EEG frames. Then, the features of temporal and the down-sampled spatial features are spliced to complete spatial-temporal information learning. In our methods, the EEG spatial correlations are effectively captured, while the temporal contexts are also learned.

The remainder of the paper is organized as follows. In section II, we describe the details of the DEAP database. In section III, we specify the details of the STILN for emotion recognition. In section IV, we conduct extensive experiments to validate the effectiveness of STILN. In section V, we discuss the performance of STILN, summarize the advantages and limitations of STILN, and compare the performance of STILN with related works. Finally, we conclude this work in section VI.

## 2. Dataset and Feature Extraction

### 2.1. Dataset

The effectiveness of STILN is examine on publicly accessible datasets, specifically DEAP for emotion analysis using physiological signals. DEAP is the multimodal emotion classification dataset consisting of EEG and peripheral physiological signals. The multi-channel neurophysiological signals of 32 healthy subjects are collected and recorded according to the 10-20 international standard of 32 leads (Fig.2). A selection of 40 1-minute-long music videos is applied to stimulate emotion. These videos are placed in 40 tracks. EEG and peripheral signals are recorded simultaneously when subjects watch a 1-minute-long music video. Trials recorded 63 seconds of EEG data from 32 electrodes at a sampling rate of 512 Hz (baseline 3 seconds, 60 seconds signal during stimulation). Subjects' emotional labels are subjectively assessed by a self-assessment model with valence and arousal dimensions.

### 2.2. EEG Preprocessing and Feature Extraction

The raw EEG is preprocessed to remove noise and improve the signal-to-noise ratio (SNR). The 32-channel raw signals are resampled at the sampling rate of 128 Hz, and band-pass-filtered in the range of 1-45 Hz to remove

EMG artifact. The independent components analysis (ICA) is conducted on the EEG data to eliminate ocular artifacts. The DEAP dataset contains 32 × 40 (subjects × trials = 1280 samples, and the amount of data are far from enough data for deep learning network training. Therefore, a 6-second window (overlapping 3-second) is used to divide the data of each trail into several segments to increase sample sizes. In addition, arousal and valence are divided into two levels, defined as high/low arousal and high/low valence, respectively. Score higher than 5 is high grade, and score lower than 5 is low grade. Thereinto, we assume that when the arousal or valence score is 5, its emotional state cannot determine, and eliminate it. Finally, 17,252 samples for arousal classification and 17,347 samples for valence classification are obtained. Thus, the problem of emotion recognition is represented as two binary classification problems of arousal and valence.

The PSD features of EEG are extracted for spatial feature learning. For each 6S-long EEG segment, PSD features are extracted in five frequency bands: Delta-band (1-4 Hz), theta-band (4-8 Hz), alpha-band (8-12 Hz), beta-band (12-20 Hz), and gamma-band (20-45 Hz). The Cartesian coordinates of 32-channel are obtained by the EEGLAB toolbox[15] (As shown in Fig.2), so that the power topography of the corresponding frequency bands can be obtained from the generated Cartesian coordinates. Electrode channels are composed into the 9×9 2D space matrices according to Cartesian coordinates, and the obtained sparse matrices is processed by biharmonic spline interpolation. To ensure learning enough edge features, the edge region of power topographic maps is zero-filled, and the 2D spatial matrices is mapped to 32×32 size. We take five different frequency bands as the channels of the power topographic maps. The generated power topographic maps of frequency bands are shown in Fig.3.

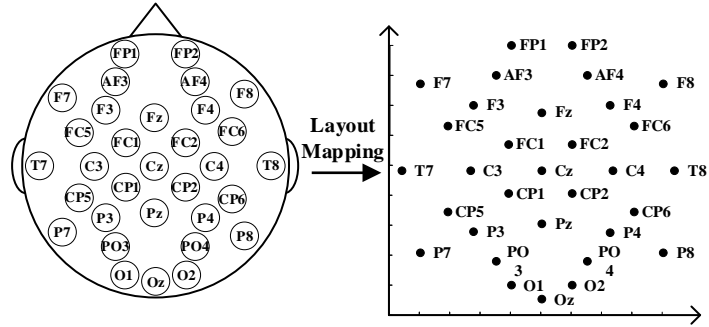

Fig.2 The cartesian coordinates of 32 electrodes based on 10–20 system.

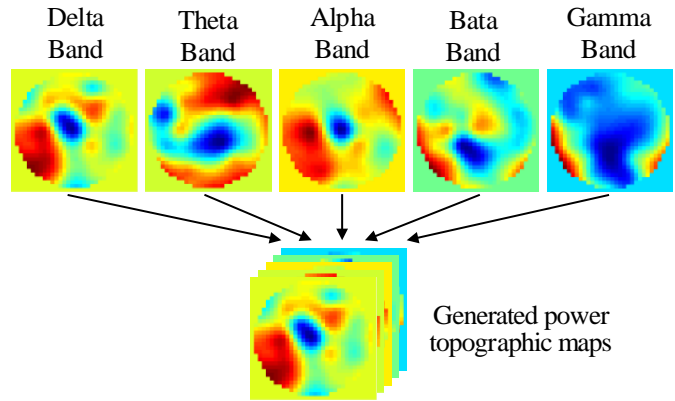

Fig.3 The generated power topographic maps from EEG.

## 3. The Novel STILN for EEG-based Emotion Recognition
### 3.1. Overview of STILN

EEG among different subjects vary significantly. For sake of maximize the accuracy for emotion recognition, researchers usually set feature extractors for specific channels or perform recalibrated operations on channels[19]-[21]. Based on this, the proposed STILN model is the method for feature learning, and introduces spatial-temporal learning and attention mechanism into the network. Firstly, the EEG spectrum fragments are combined with the electrode position to form the power topographic maps. Then the training samples are sent to STILN for training, and the ADAM optimizer updates the corresponding network parameters. Finally, the trained model is utilized to classify the pre-processed test samples. The configuration of the STILN model is shown in Fig.4, and the parameters are shown in Table 1.

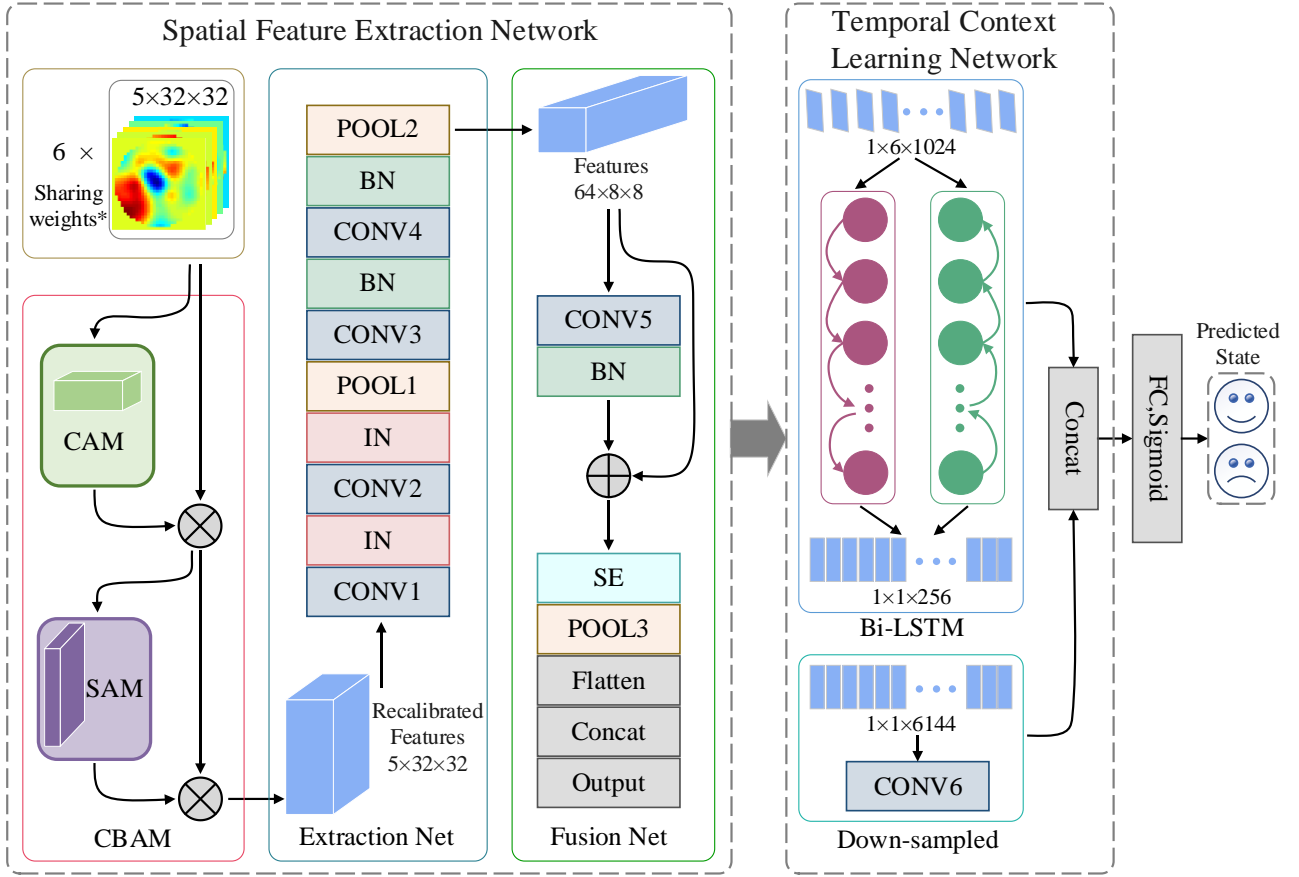

Fig.4 Overview of STILN. Note: Sharing weights* denotes that six EEG 1-second segments share weights during spatial feature extraction network training. ⊗ denotes that the multiplication of the tensors; ⊕ denotes that the addition of the tensors; CONCAT denotes that splicing of the tensor; FC denotes that the full connection layer; Detailed structure of channel attention module (CAM) and spatial attention module (SAM) show in fig.5; Tensors of input Bi-LSTM and CONV6 are spliced according to different dimensions.

Table 1 Parameters of STILN

| Layer | Operation | Kernel size/stride | Activation function |
|---|---|---|---|
| CBAM | Channel and spatial-wise attention | 7×7/1 | ReLU |
| CONV1 | Convolutional(2D) | 5×5/1 | ReLU |
| CONV2 | Convolutional(2D) | 5×5/1 | ReLU |
| CONV3 | Convolutional(2D) | 3×3/1 | ReLU |
| CONV4 | Convolutional(2D) | 3×3/1 | ReLU |
| CONV5 | Convolutional(2D) | 3×3/1 | ReLU |
| SE | Channel-wise attention | 1×1/1 | ReLU/ Sigmoid |
| Bi-LSTM | LSTM | - | ReLU |
| CONV6 | Convolutional(1D) | 1×1/48 | ReLU |
| FC | Linear | - | ReLU |
| Output | Linear | - | Sigmoid |

*3.2. Spatial Feature Extraction Network*

According to the contribution of different electrodes and frequency bands to the classification, the corresponding weights are recalibrated by CBAM[22] to improve the classification accuracy. After the power topographic maps input the CNN-based spatial feature extraction network, CBAM continuously calculates the feature attention maps along the channel and spatial. Finally, the recalibrated power topographic maps are obtained by multiplying the elements of the attention maps and the input power maps. We assume that the power topographic maps of the input feature extraction network is $F \in \mathbb{R}^{H \times W \times C}$, $H$ and $W$ represent the size of the feature map, and $C$ represents the number of feature map channels. Therefore, the recalibrating operation process of CBAM can be expressed as follows:

$$F_{CAM} = M_C(F) \otimes F$$

$$F_{CBAM} = M_S(F_{CAM}) \otimes F_{CAM}$$

where, $F_{CAM}$ represents the output of channel weights recalibrating module in CBAM, $F_{CAM} \in \mathbb{R}^{H \times W \times C}$, $M_C(\cdot)$ represents the operation of channel weights recalibrating module. $F_{CBAM}$ represents the output weights recalibrated by spatial module, that is, the final output of CBAM. $F_{CBAM} \in \mathbb{R}^{H \times W \times C}$. $M_S(\cdot)$ represents the operation of the spatial weight recalibrating module. $\otimes$ represents the element-level multiplication of feature map channels. Of these, $M_C(\cdot)$ and $M_C(\cdot)$ are calculated as follows:

$$M_C(F) = \sigma(MLP[G_{Avg}(F)] + MLP[G_{Max}(F)])$$

$$M_S(F) = \sigma(G_{CONV}(concat([G_{Avg}(F)], [G_{Max}(F)])))$$

In the above formula, $\sigma(\cdot)$ denotes the sigmoid function, *MLP* denotes the operation of multi-layer perceptron, $G_{Avg}$ denotes average pooling operation, $G_{Max}$ denotes the operation of maximum pooling, $G_{CONV}$ denotes the convolution operation with convolution kernel of 7×7, *concat* denotes the splicing operation of tensors.

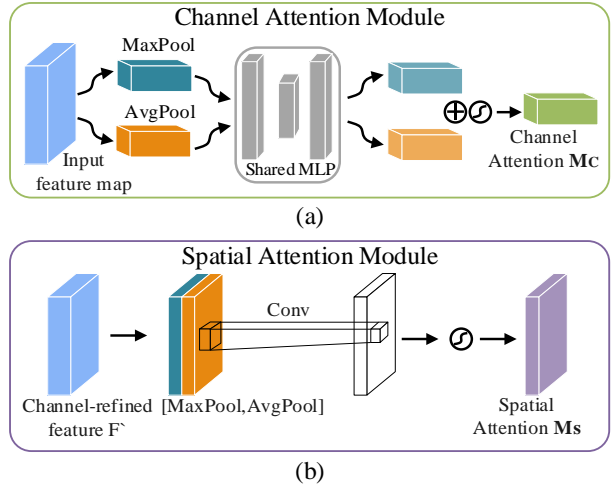

Fig.5 CBAM structure. (a) is channel attention module; (b) is spatial attention module. Note: MaxPool and AvgPool indicates maximum pooling operation and average pooling operation respectively; MLP denotes that Multilayer Perceptron; Conv denotes the 2D convolution, and kernel size is 7×7.

One of the difficulties of emotion recognition is solving the inconsistency of feature distribution caused by individual differences in EEG. Instance normalizations (INs) are confirmed to be successful in transfer learning tasks[23]. In EEG emotion recognition, INs independently normalize feature channels to alleviate the differences caused by various EEG between individuals. In parallel, Batch Normalizations (BNs) significantly improve the performance of classification tasks, and the combination of INs and BNs may achieve satisfactory results[24]. Therefore, we design EEG feature extraction network combining INs and BNs. The specific network structure refers to Fig.4 and Table 1. The output of the feature extraction network is $F_{SFE}$, $F_{SFE} \in \mathbb{R}^{H \times W \times C}$.

The deep features of EEG have outstanding effects on emotion recognition. Inspired by the residual network model structure[25], a feature fusion block suitable for EEG is proposed, as shown in Fig. 4. The input of feature maps and the output of convolution are addition performed at the element level, and the output of the feature fusion

block is activated by the ReLU function. The formula operations for this layer are expressed as follows:

$$F_{Res} = \rho(F_{SFE} \oplus [CONV2D(F_{SFE})])$$

where, $F_{Res}$ represents the outputs of deep feature fusion block, $F_{Res} \in \mathbb{R}^{H \times W \times C}$. $\sigma(\cdot)$ is the ReLU activation function. $CONV2D$ is convolution operation, where the convolution kernel size is 3×3 and the number of convolution kernels is 64. $\oplus$ represents the addition operation of the corresponding channels.

We propose a strategy of twice recalibrating the feature map channel weights by the squeeze and excitation (SE) module [26]. SE module recalibrates the weights of features in different channels through adaptive calibration of the interdependence of EEG. The adopted SE module includes two main operation operations: squeeze and excitation. The output of SE module is the proportional operation between the excitation scalar and the input network features mapping. Suppose $u_c \in \mathbb{R}^{H \times W \times C}$ is the feature mapping tensor of $F_{Res}$ outputted by deep feature fusion layer, $U = [u_1, u_2, ..., u_c]$. The operation process of SE module is as follows:

$$S_C = G_{sq}(u_c) = \frac{1}{H \times W} \sum_{i=1}^{H} \sum_{j=1}^{W} u_c(i,j)$$

$$E = G_{ex}(S, W) = \sigma(W_2 \rho(W_1 S))$$

$$\grave{F_{SE}} = G_{scale}(E, U)$$

where, $S_C$ denote output of channels in the squeeze operation, $S = [S_1, S_2, ..., S_C]$, $S \in \mathbb{R}^c$ denote the descriptor of C channels in the feature map. $G_{sq}(\cdot)$ represents the squeeze operation. $E$ represents output of channels in the excitation operation, and $G_{ex}(\cdot)$ represents the excitation operation. $W_1 \in R^{\frac{C}{r} \times C}$ and $W_2 \in R^{C \times \frac{C}{r}}$ are the weights of ReLU activation function $\rho(\cdot)$ and sigmoid function $\sigma(\cdot)$ respectively. $r$ is the dimension reduction ratio, which is set to 4. $G_{scale}(\cdot)$ represents the multiplication of channel elements between scalar $E$ and feature map $U$. $\grave{F_{SE}} \in R^{H \times W \times C}$ represents output of the feature map after twice recalibrating.

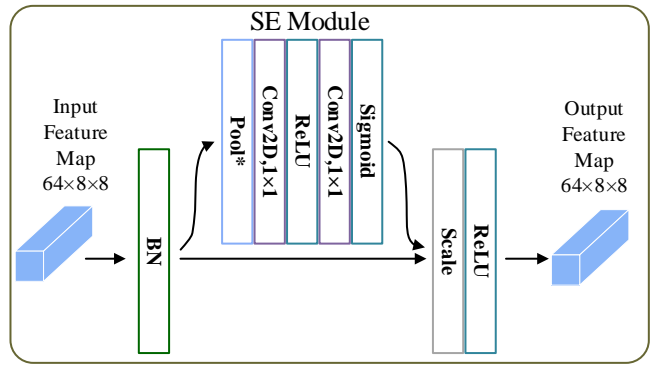

Fig.6 The SE Module structure. Note: The pool∗ denotes the global average pooling.

### 3.3. Temporal Contexts Learning Network

LSTM is a type of recurrent neural network (RNN), and remembers long sequences of input data[27]. LSTM captures remote dependencies through gate control and storage units, but it only utilizes information from previous time steps. Bi-LSTM captures the temporal contexts of the sequence through bidirectional learning[28]. Therefore, we utilize Bi-LSTM to learn temporal contexts.

In the spatial feature extraction network, we input the power topographic maps of 1-second EEG segment in trials (each sample is divided into 6 segments) into the network. And in temporal information learning, in order to learn the EEG features in temporal contexts, 6-separate spatial feature tensors are connected to obtain EEG time series containing spatial features. Assuming that $F_{CAT} \in R^{H \times W \times C}$ represents the vector after the 6-seconds spatial feature map is connected, the operation process of Bi-LSTM is as follows:

$$F_{mutual} = concat[L_{left}(F_{CAT}), L_{right}(F_{CAT})]$$

where, $L(\cdot)$ represents the unit operation in Bi-LSTM, $L_{left}(\cdot)$ represents the operation in the left direction, $L_{right}(\cdot)$ represents the operation in the right direction, and $F_{mutual} \in \mathbb{R}^{2d \times N}$ represents the spatial features of EEG obtained by Bi-LSTM, where $d$ is the number of hidden layers in Bi-LSTM. Finally, the down-sampled spatial features are linked with the temporal contexts to complete the co-learning of spatial-temporal information. The obtained spatial-temporal features are fed into the fully connected layer, and applied to arousal or valence prediction after the sigmoid function operation.

## 4. Experiments
*4.1 Experimental Setups*

The experiment is performed on the public DEAP dataset. The leave-one-out cross-validation (LOOCV) method is used to evaluate the classification performance of the STILN model. Specifically, the samples of one subject as the test dataset and the samples of the other 31 subjects as the training dataset, until samples of all 32 subjects are set to the test set once. The arousal and valence SAM scores 1-4 are divided into low grades and 6-9 are divided into high grades. The arousal samples are divided into 7144 low arousal samples and 10108 high arousal samples. The valence samples are divided into 6973 low valence samples and 10374 high valence samples. as shown in Table 2.

Table 2 The data labeling scheme and number of samples per emotion class

| task | status | Emotion class | Label scores | Data numbers |
|---|---|---|---|---|
| 2-class classification | Arousal | LA | A<5 | 7144 |
| | | HA | A>5 | 10108 |
| | Valence | LV | V<5 | 6973 |
| | | HV | V>5 | 10374 |

Note: LA denotes low arousal; HA denotes high arousal; LV denotes low valence; HV denotes high valence.

Configuration parameters are required for running the STILN model. ADAM is selected as the parameter optimizer. The learning rate is set to 0.0005, the batch size is set to 256. The cross-entropy loss selected as loss function, and the emotional labels of the samples are processed by one-hot coding. All networks in this work are implemented in Pytorch framework with NVIDIA GeForce RTX 2060 SUPER GPU. In addition, the accuracy ($P_{acc}$) and F1 score ($P_{F1}$) are set for model evaluation, the distribution of test results is reflected by standard deviation.

*4.2. Experimental Results*

Table 3 Average accuracy and F1 scores of subject-independent EEG-based experiments on the DEAP database

| Strategies | Arousal | | Valence | |
|---|---|---|---|---|
| | $P_{acc}$ | $P_{F1}$ | $P_{acc}$ | $P_{F1}$ |
| TOP10 Subjects | 0.8132 (0.0602) | 0.7744 (0.0776) | 0.7747 (0.0331) | 0.7776 (0.0301) |
| ALL Subjects | 0.6831 (0.1124) | 0.6826 (0.0838) | 0.6752 (0.0873) | 0.68 (0.0969) |

Note: Standard deviation in the brackets.

The subject-independent experimental classification results of arousal and valence level are shown in Table 3. The arousal and valence accuracy of the top ten subjects were 0.8132 and 0.7747, and the F1 scores are 0.7744 and 0.7776, respectively. The arousal and valence accuracy of all 32 subjects are 0.6831 and 0.6752, respectively, and the F1 scores are 0.6826 and 0.68, respectively. In addition, Fig.7 shows the classification accuracy and F1 score histogram of 32 subjects. Experimental results indicate that the proposed EEG feature extraction network achieves outstanding performance in subject-independent arousal and valence classification.

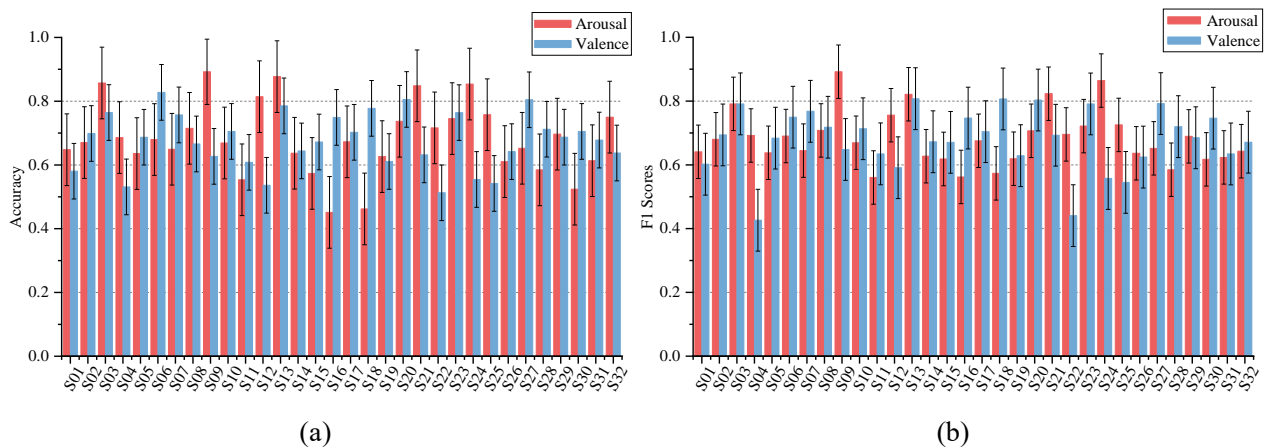

(a)                            (b)

Fig7. The classification accuracy and F1 scores histogram of subjects. (a) denotes accuracy histogram of each subject, (b) denotes F1 score histogram of each subject. Note: Abscissa shows the subject number.

Since the parameters of the deep learning model have significant impact on its performance, we study the different model settings and select the parameters with the best performance. In this work, we hold that the number of hidden layers of Bi-LSTM and the learning rate of the optimizer have great effect on the performance. Therefore, the two parameters for different values are set to be tested. For the hidden layer of Bi-LSTM, five comparison parameters of 16,32,64,128,256 are set, and three parameters of 0.0001,0.0005 and 0.001 are set for the comparison of learning rate. In the parameter comparison experiment, except for the parameters to be compared, other network configurations are the same.

The effects of different numbers of hidden layers and different learning rates on the performance of accuracy are shown in Table 4 and 5, respectively. According to the results of classification in Table 4, when the number of hidden layers is 64, STILN achieves the highest average performance of accuracy (arousal is 0.6831, valence is 0.6752) and F1 score (arousal is 0.6826, valence is 0.68) in arousal and valence classification. According to the box plot of the accuracy and F1 score of each subject under different parameters in Fig.8, it can be analyzed that the performance index distribution of Bi-LSTM is more concentrate in the 64-layer hidden layer state, and the proposed emotion recognition network is also more robust. When the number of hidden layers is less, the features required for classification cannot be fully learned in the model, but the number of hidden layers is large, the model appears overfitting. According to Table 5 and Fig.9, the learning rate is set to 0.0005, the proposed model achieves the best performance, results indicate that too long or too short learning steps cannot effectively learn the necessary EEG features.

Table 4 Average accuracy and F1 score in EEG-based emotion recognition for different hidden layers of Bi-LSTM

| hidden layers | Arousal | | Valence | |
|---|---|---|---|---|
| | $P_{acc}$ | $P_{F1}$ | $P_{acc}$ | $P_{F1}$ |
| 16 | 0.6631 (0.1247) | 0.6602 (0.0977) | 0.6698 (0.0974) | 0.6682 (0.0958) |
| 32 | 0.6496 (0.1293) | 0.6491 (0.0926) | 0.6599 (0.1065) | 0.6606 (0.1051) |
| **64** | **0.6831 (0.1124)** | **0.6826 (0.0838)** | **0.6752 (0.0873)** | **0.68 (0.0969)** |
| 128 | 0.661 (0.143) | 0.6576 (0.1092) | 0.6672 (0.0989) | 0.6542 (0.1128) |
| 256 | 0.6696 (0.1292) | 0.6621 (0.1107) | 0.6663 (0.1101) | 0.6612 (0.1105) |

Note: Standard deviation in the brackets.

Table 5 Average accuracy and F1 score in EEG-based emotion recognition of the model with different learning rates

| Learning Rate | Arousal | | Valence | |
|---|---|---|---|---|
| | $P_{acc}$ | $P_{F1}$ | $P_{acc}$ | $P_{F1}$ |
| 0.0001 | 0.6618 (0.1168) | 0.6545 (0.1046) | 0.6615 (0.1046) | 0.6621 (0.1043) |
| **0.0005** | **0.6831 (0.1124)** | **0.6826 (0.0838)** | **0.6752 (0.0873)** | **0.68 (0.0969)** |
| 0.001 | 0.6476 (0.1451) | 0.6578 (0.1009) | 0.6508 (0.103) | 0.6581 (0.1004) |

Note: Standard deviation in the brackets.

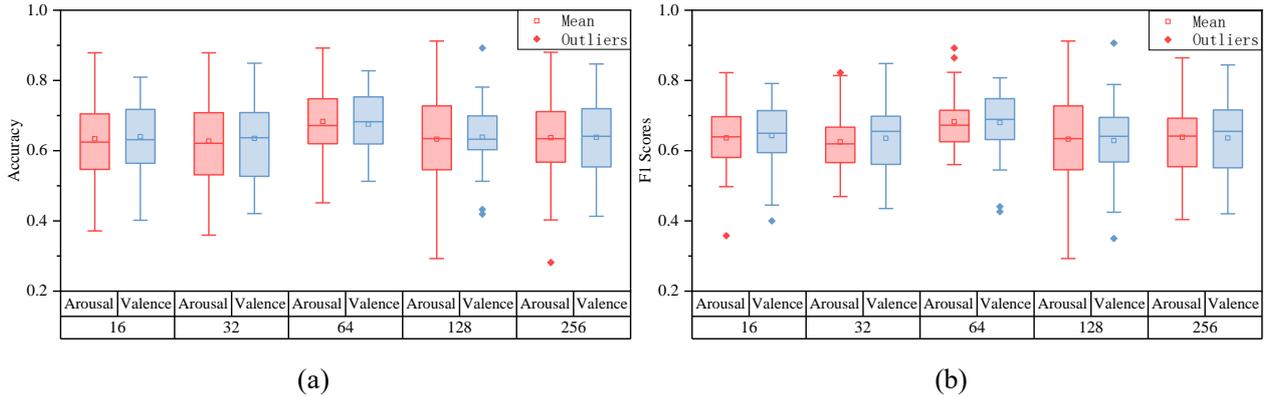

(a)      (b)

Fig.8 Box-plot of accuracy and F1 scores for 32 subjects in different hidden layers. (a) denotes Box-plot of accuracy, (b) denotes Box-plot of F1 scores. Note: The number at the bottom of the horizontal coordinate shows the hidden layers number.

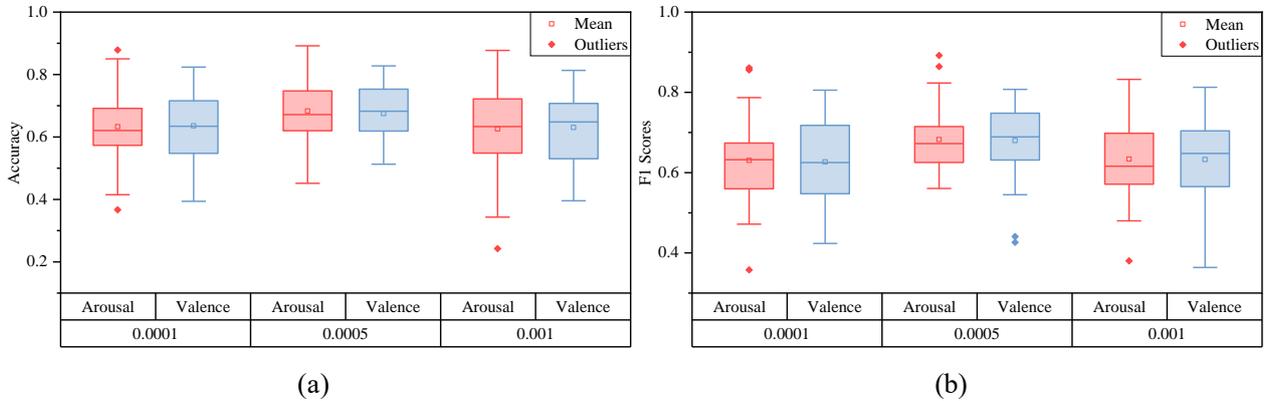

(a)      (b)

Fig.9 Box-plot of accuracy and F1 scores for 32 subjects in different learning rate. (a) denotes Box-plot of accuracy, and (b) denotes Box-plot of F1 scores. Note: The number at the bottom of the horizontal coordinate shows the learning rate number.

    In this work, we compare STILN with classical deep neural networks, and test the improvement of performance for the proposed mothed compare to the common methods. Table 6 shows the performance of classification for different methods. Compared with CNN, LSTM, and DBN, the performance of accuracy for arousal in our proposed method is improved by 7.59 %, 9.92 %, and 8.49 % respectively, and the performance of accuracy for valence is improved by 6.09 %, 8.87 %, and 11.54 % respectively. Besides, we also reimplement the latest spatial and temporal EEG encoding networks, CNN-LSTM and DenseNet. Compared with these two networks, our method has 6.21 % and 4.94 % improvement in the accuracy of arousal, and 3.66 % and 2.95 % improvement in the accuracy of valence. Collectively, STILN achieves more stable performance on the DEAP standard dataset. The comparison results verify the effectiveness of the proposed STILN model in emotion recognition.

Table 6 The results of deep networks on the DEAP database

| Methods | Arousal | | Valence | |
|---|---|---|---|---|
| | $P_{acc}$ | $P_{F1}$ | $P_{acc}$ | $P_{F1}$ |
| CNN[11] | 0.6072 (0.101) | 0.5888 (0.1083) | 0.6143 (0.086) | 0.6068 (0.0808) |
| LSTM[29] | 0.5839 (0.0904) | 0.5261 (0.0985) | 0.5865 (0.0816) | 0.5331 (0.089) |
| DBN[30] | 0.5982 (0.0872) | 0.5749 (0.0823) | 0.607 (0.0893) | 0.5972 (0.0841) |
| CNN-LSTM [31] | 0.621 (0.0963) | 0.6047 (0.0999) | 0.6386 (0.0855) | 0.6136 (0.0737) |
| DenseNet [32] | 0.6337 (0.0705) | 0.615 (0.0747) | 0.6457 (0.0748) | 0.6258 (0.0706) |
| **STILN (ours)** | **0.6831 (0.1124)** | **0.6826 (0.0838)** | **0.6752 (0.0873)** | **0.68 (0.0969)** |

Note: Standard deviation in the brackets.

### *4.3. Ablation Experiment*

To evaluate the function of proposed modules in the model, we perform ablation experiments. Ablation experiment steps for modules are based on the LOOCV experiments design. Fig.10(a) shows the STILN principal network structure with the complete module, named NET0. To verify the performance of the CBAM attention mechanism, the NET1 structure shown in Fig.10(b) is designed. INs of the CNN-based spatial feature learning module are replaced by BNs to verify the validity, and the designed network structure NET2 is shown in Fig.10(c). In Fig.(d), NET3 is designed to verify the deep feature fusion, using a layer of 2D convolution to replace the proposed fusion structure. The verification of performance for the SE module is implemented by the NET4 architecture, as shown in Fig.10(e). Bi-LSTM for temporal contexts features is bidirectional temporal information learning. To verify the performance of Bi-LSTM, we replace Bi-LSTM with LSTM in NET6. The detailed structure is shown in Fig.10(f).

Table 7 shows the performance of arousal and valence for ablation experiment. The complete STILN (NET0) improves the accuracy of arousal by 0.74 % -3.49 % over other structures and F1 scores by 1.87 % -3.48 %. In the classification of valence, compared with other ablation experimental structures, the accuracy is increase by 0.56 % -2.99 % in NET0 and its F1 score increase by 1.52 % -4.44 %. Compared with other variant structures, the performance of complete STILN has been significantly improved. Ablation experiment verifies the important role of CBAM attention mechanism, INs, deep fusion structure, SE module, and Bi-LSTM in the proposed EEG emotion classification model.

Table 7 The results of different network configurations on the DEAP datasets in ablation experiments

| NET | Arousal | | Valence | |
|---|---|---|---|---|
| | $P_{acc}$ | $P_{F1}$ | $P_{acc}$ | $P_{F1}$ |
| **NET0** | **0.6831 (0.1124)** | **0.6826 (0.0838)** | **0.6752 (0.0873)** | **0.68 (0.0969)** |
| NET1 | 0.654 (0.1056) | 0.6477 (0.0963) | 0.6452 (0.1027) | 0.6355 (0.1266) |
| NET2 | 0.6756 (0.1549) | 0.6638 (0.1278) | 0.6511 (0.1115) | 0.6623 (0.1019) |
| NET3 | 0.6481 (0.1427) | 0.651 (0.118) | 0.6485 (0.1021) | 0.6579 (0.0835) |
| NET4 | 0.6619 (0.1313) | 0.6478 (0.1122) | 0.6603 (0.1041) | 0.6647 (0.1021) |
| NET5 | 0.6677 (0.1213) | 0.6575 (0.1117) | 0.6695 (0.1005) | 0.6499 (0.1006) |

Note: Standard deviation in the brackets.

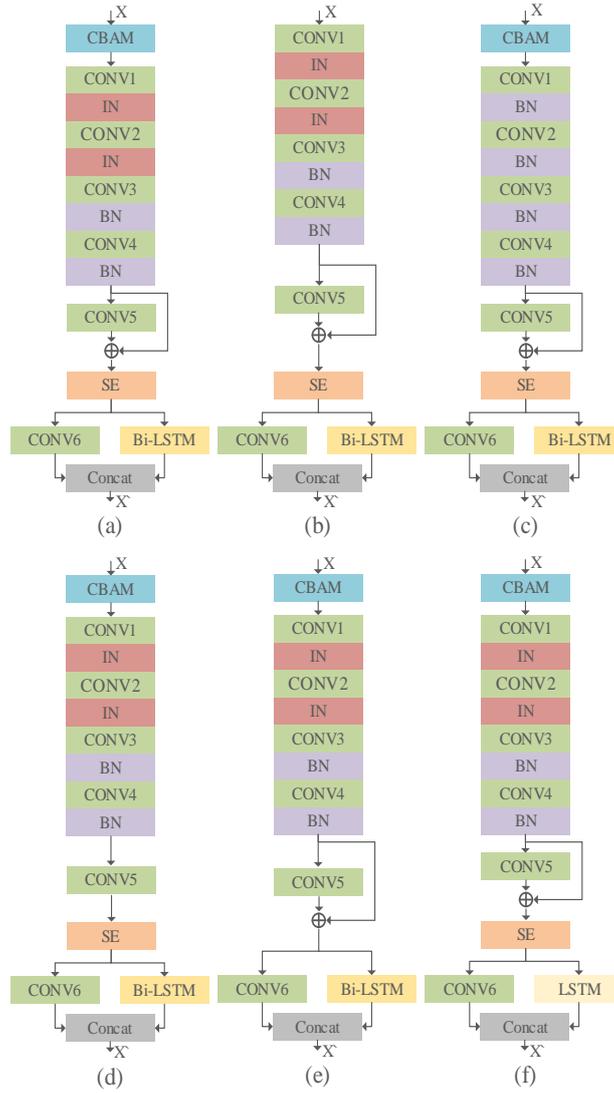

Fig.10 The network structures for ablation experiment. (a): NET0; (b): NET1; (c): NET2; (d): NET3; (e): NET4; (f): NET5. Note: The network parameters are shown in the table1. LSTM uses the same hidden layer as Bi-LSTM.

## 5 Discussion

### 5.1. Advantages of STILN

In our proposed model, the power topographic maps about electrodes and different frequency bands are constructed, and it focus on the links between different electrodes and adjacent frequency bands. The CBAM and SE channel attention mechanism are utilized to realize channel and spatial recalibrated processing with different contributions to classification. The individual difference of EEG between subjects is the obstruction of emotion recognition. INs effectively reduce the difference between individuals. After completing spatial feature extraction, the temporal contexts are learned by Bi-LSTM while the critical EEG frames are emphasized. Then, the down-sampled spatial features are linked with the temporal contexts to complete the co-learning of spatial-temporal information. The accuracy of emotion recognition is improved.

The proposed STILN model achieves higher classification performance for emotion recognition (arousal:0.6831, valence:0.6752). As shown in Table 3, the standard deviation of the average classification accuracy in our proposed model is smaller, results indicate that the EEG features of individual differences are concerned in the method. In Table 4 and Table 5, we analyze affected with different numbers of hidden layers of Bi-LSTM and the learning rate of the training network on the performance of classification. When the number of hidden layers is set to 64 and the learning rate is set to 0.0005, the proposed model achieved optimal performance. We also contrast the performance of proposed model with the classic networks of classification for emotion recognition. The results in Table 6 shows that the classification performance of our model is significantly improved compared with classical methods. Finally, we conduct ablation experiment on the performance of the modules. The results prove that proposed modules in the model possess an irreplaceable role in emotion recognition. Taken together, the STILN model shows outstanding performance in related research, and the proposed model is effective for emotion recognition.

*5.2. Comparison with Related Research*

We compare the performance of STILN with the research schemes in the past four years on the DEAP dataset. Table 8 lists the performance of classification for comparison schemes. Except for references[33] and [38] using the K-fold cross-validation method, the remaining research uses LOOCV. Comparing results, the proposed STILN possesses higher accuracy of arousal and valence than the performance of networks in [33]-[40] (except[38]). Our method is more accurate in classification of arousal, but the performance of classification for valence is not as better as [38]. Through analysis, [38] adopts K-fold cross-validation, and the EEG of the same subjects is distributed in the training set and the testing set, and resulting in higher accuracy for test. Furthermore, we find that the methods using PSD features of EEG superior to the methods applying other EEG features (differential entropy, wavelet entropy, etc.). Overall, our STILN model achieves relatively outstanding performance of classification.

In parallel, the limitation of STILN is also worth exploring. As in all deep-learning methods, the STILN model requires a significant amount of data for training. In this study, the usage of a 6-second sliding overlap window is to obtain a larger data (17,252 samples of classification for arousal, and 17,347 samples of classification for valence). However, it still does not reach the size of samples for training our deep network, so the performance of STILN is restricted. Moreover, the analysis of standard deviation for accuracy and F1 score demonstrates that the results of classification are unevenly distributed. The reason maybe is that the network is insensitive to EEG with the extreme individual differences.

Table 8 Performance comparison with related works

| Related researches | Features | Classifier | Validation method | Accuracy ($P_{acc}$) |
|---|---|---|---|---|
| Chen *et al.* (2019)[33] | PSD | GRU | 10-folds cross-validation | Arousal:0.6790 Valence:0.6650 |
| Zhong *et al.* (2020)[34] | DE Features | SE_CNN | Leave-one-out Cross-Validation | Arousal:0.6623 Valence:0.6850 |
| Yin *et al.* (2020)[35] | Frequency Spectral | LSSVM | Leave-one-out Cross-Validation | Arousal:0.6510 Valence:0.6797 |
| Zhang *et al.* (2020)[36] | Temporal-Frequency Spectral | SSFE | Leave-one-out Cross-Validation | Arousal:0.6521 Valence:0.6635 |
| Liang *et al.* (2021)[37] | Temporal Domain Features | EEGFuseNet | Leave-one-out Cross-Validation | Arousal:0.5855 Valence:0.5644 |
| Yang *et al.* (2022)[38] | Sample entropy and Wavelet entropy | SVM | K-folds cross-validation | Arousal:0.6420 **Valence:0.7010** |
| He *et al.* (2022)[39] | Temporal Domain Features | TSNs+ADDA | Leave-one-out Cross-Validation | Arousal:0.6433 Valence:0.6325 |
| Wang *et al.* (2022)[40] | PSD | HSLT | Leave-one-out Cross-Validation | Arousal:0.6575 Valence:0.6663 |
| **The proposed STILN(ours)** | PSD | STILN | Leave-one-out Cross-Validation | **Arousal:0.6831** Valence:0.6752 |

Note: DE denotes Differential entropy. GRU denotes Gated Recurrent Neural Network; SE_CNN denotes Convolutional Neural Network for adding channel attention; LSSVM denotes Least Squares Support Vector Machine; SSFE denotes the shared-subspace feature elimination approach; EEGFuseNet denotes a practical hybrid unsupervised deep convolutional recurrent generative adversarial network; SVM denotes Support Vector Machine;

TSNs+ADDA denotes the feasibility of combining temporal convolutional networks (TCNs) and adversarial discriminative domain adaptation (ADDA) algorithms; HSLT denotes Hierarchical Spatial Learning Transformer.

## 6. Conclusion

In this work, we propose a spatial-temporal information learning network for EEG-based emotion recognition. The power topographic maps generated according to PSD features are adopted to evaluate the STILN performance. The EEG spatial correlations and temporal contexts are effectively learned and fused, and it greatly enhances the accuracy of emotion recognition. The STILN has achieve anticipated results in subject-independent experiment with the accuracy of arousal and valence 0.6831/0.6752 in the DEAP database respectively. Experimental results indicate that the performance of STILN is outstanding than the previous methods. In addition, we also analysis the limitation of STILN. In the future work, the STILN model performs performance verification on more benchmark datasets. Further, we will proceed to settle the problem of larger individual differences in EEG, and explore multimodal data for cross-subject emotion recognition.